# Quark Matter 2004 – Oakland Ca – Poster Presentation

## Excitation Function of $<p_t>$ and Net Charge Fluctuations at RHIC

*Claude A Pruneau*
*(for the Star Collaboration)*

Physics and Astronomy Department
Wayne State University
666 West Hancock,, Detroit,
MI 48201   USA
(Dated: January 11, 2004)

We present measurements of net charge and transverse momentum fluctuations in Au + Au collisions at $\sqrt{s_{NN}}$ = 20, 130, and 200 GeV reported in our QM 2004 poster using measures $\nu_{+-,dyn}$ and $<\Delta p_{t,1} \Delta p_{t,2}>$. We observe the dynamical fluctuations are finite at all three energies and exhibit a rather modest dependence on beam energy. We also find both net charge and $p_t$ fluctuations violate the trivial 1/N scaling expected for nuclear collisions consisting of independent nucleon-nucleon interactions. We speculate the observed centrality dependence arises due to radial flow and might provide insight on the width and shape of the radial flow profile.

I.    Introduction

Anomalous transverse momentum and net charge event-by-event fluctuations were proposed as indicators of the formation of quark gluon plasma (QGP) in the midst of high-energy heavy ion collisions. Measurements performed by the STAR and PHENIX collaborations showed transverse moment and net charge dynamic fluctuations in Au + Au at $\sqrt{s_{NN}}$ = 130 GeV are definitely finite but small relative to predictions invoking 1st order phase transition [1,2,3,4]. The magnitude of the fluctuations was found to be in qualitative agreement with HIJING predictions although the data exhibit finite centrality dependence not found in the HIJING calculations.  Although the net charge and $p_t$ dynamic fluctuations measured at $\sqrt{s_{NN}}$ = 130 GeV are not as large as anticipated, there remains the possibility that the observed fluctuations are reduced from their expected value due to some final state effects, or because only a small fraction of the system actually produces a QGP.  One is thus lead to wonder whether the fluctuations may be found to vary with beam energy thereby indicating the production of a QGP above a critical threshold, or with progressively increasing probability at higher energies. In this paper, we consider this possibility by investigating how the strength of the net charge and $p_t$ dynamic fluctuations vary with beam energy in Au + Au collisions ranging in center of mass energy from the highest SPS energy to the highest RHIC energy, and relative to p + p collisions at $\sqrt{s_{NN}}$ = 130 GeV.

The experimental method is described in Section II. Results are presented in Sections III and IV for net charge and p$_t$ fluctuations respectively. Section V presents a summary of the results and conclusions of this work.

## II. Experimental Method

The data used in this analysis were measured using the Solenoid Tracker at RHIC (STAR) detector during the 2001, and 2002 data RHIC runs at Brookhaven National Laboratory. They include Au + Au collisions data collected at $\sqrt{s_{NN}}$ = 20, 130, and 200 GeV as well as p + p collision data measured at $\sqrt{s_{NN}}$ = 200 GeV. The $\sqrt{s_{NN}}$ = 130, and 200 data were acquired with minimum bias triggers accomplished by requiring a coincidence of two Zero Degree Calorimeters (ZDCs) [5] located 18 m from the center of the interaction region on either sides of the STAR detector. For 20 GeV data, a combination of minimum bias and central triggers was used. The centrality trigger was achieved using a set of scintillation detectors surrounding the main TPC (Central Trigger Barrel).

The analysis is based on charged particle track reconstruction measurements performed with the STAR main Time Projection Chamber (TPC) [6]. The TPC is located in a large solenoidal magnetic field producing a uniform axial magnetic field. The magnetic field was set to 0.25 T for the 20 and 130 GeV data, and 0.5 T for the 200 GeV data. The net charge fluctuations analysis was conducted with tracks from the TPC with transverse momentum in the range 0.2 GeV/c < p$_t$ < 5.0 GeV with pseudorapidity |η| < 1.0 whereas the p$_t$ fluctuations analysis used a somewhat more restrictive momentum cut of 0.2 GeV/c < p$_t$ < 2.0 GeV. In order to limit the net charge fluctuations analysis to primary tracks only (i.e. particles produced by the collision), we selected tracks that pass within a distance of closest approach (DCA) to the collision vertex of 3 cm. Reducing the cut to 1 cm leads to a systematic increase of measured $\nu_{+-,dyn}$ (introduced and defined below) by less than 5 %. The DCA cut was set to 1 cm in the p$_t$ fluctuations analysis. An additional quality cut, requiring tracks consist of at least 20 measured TPC hits, was used to limit split track contamination.

Events were selected for analysis if their collision vertex lied within a maximum distance from the center of the TPC and they passed a minimal track multiplicity cut. The maximal distance in the plane transverse to the beam direction was to 1 cm. The maximal distance along the beam axis was set respectively to 75 cm for the 20, and 130 GeV data, and further restricted to 25 cm for the 200 GeV Au + Au data. A maximum of 75 cm was used for the p + p data. The wide 75 cm cut was used to maximize the event sample used in this analysis. We found that within statistical errors, observables measured in this analysis are insensitive to shorter cut distances but finitely diverged for larger values. The minimal track multiplicity cut was set to 3 for Au + Au collisions and 0 for p + p collisions.

The collision centrality is inferred and binned according to the multiplicity of all charged particles measured in the TPC within |η|<0.5. The centrality bins were calculated as a fraction of this multiplicity distribution starting at the highest multiplicities. The ranges used were 0-5% (most central collisions), 5-10%, 10-20%, 20-30%, 30-40%, 40-50%, 50-60%, 60-70%, and 70-80% (most peripheral). Each centrality bin is associated with an average number of participating nucleons, $N_{part}$, using a Glauber Monte Carlo calculation[7].

Our study of dynamical net charge fluctuations dependence on the beam energy is based on the observable $\nu_{+-,dyn}$ used in the first STAR measurement [1] of net charge fluctuations in Au + Au collisions at $\sqrt{s_{NN}}$ = 130 GeV. The definition of $\nu_{+-,dyn}$, its properties, and relationships to other measures of event-by-event net charge fluctuations were motivated and presented in details in Ref. [8]. The robustness of $\nu_{+-,dyn}$ as an experimental observable was nicely discussed on the basis of Monte Carlo toy models by Nystrand *et al.* [9]: $\nu_{+-,dyn}$ is insensitive to the details of the detector response and efficiency. It is calculated as the difference between the relative multiplicity difference $\nu_{+-}$ defined as

$$\nu_{+-} = \left\langle \left( \frac{N_+}{\langle N_+ \rangle} - \frac{N_-}{\langle N_- \rangle} \right)^2 \right\rangle$$

and its statistical limit, $\nu_{+-,stat}$ found to be:

$$\nu_{+-,stat} = \frac{1}{\langle N_+ \rangle} + \frac{1}{\langle N_- \rangle}$$

From a theoretical standpoint, $\nu_{+-,dyn}$ can be expressed in terms of two-particle integral correlation functions as follows.

$$\nu_{+-,dyn} = R_{++} + R_{--} - 2R_{+-}$$

The analysis of transverse momentum fluctuations is conducted, in a similar fashion, on the basis of the average momentum correlation $<\Delta p_{t,1} \Delta p_{t,2}>$ defined as

$$\langle \Delta p_{t,1} \Delta p_{t,2} \rangle = \frac{1}{N_{event}} \sum_{k=1}^{N_{event}} \sum_{j=1, i \neq j}^{N_k} \frac{\left( p_{t,j} - \langle\langle p_t \rangle\rangle \right)\left( p_{t,i} - \langle\langle p_t \rangle\rangle \right)}{N_k (N_k - 1)}$$

where the double brackets denote the momentum average defined as

$$\langle\langle p_t \rangle\rangle = \frac{1}{N_{event}} \sum_{k=1}^{N_{event}} \langle p_t \rangle_k$$

and $N_{event}$ is the number of events analyzed, $N_k$, the number of particles from event "k", $<p_t>_i$, the average $p_t$ of particles of the i$^{th}$ event, and $p_{t,j}$ the transverse momentum of the i$^{th}$ particle in an event.

In principle, both $\nu_{+-,dyn}$ and $<\Delta p_{t,1} \Delta p_{t,2}>$ can be evaluated for varied ranges of pseudorapidity and azimuthal coverage. In this presentation, we focus mainly on data measured within the pseudorapidity range |η|<0.5, although we also discuss in the next section a measurement of the relative strength of the net charge fluctuations as function of the integrated pseudorapidity range for two bins of collision centrality.

### III. Net Charge Fluctuation Results

STAR reported that the dynamical fluctuations $\nu_{+-,dyn}$, in Au + Au collisions at $\sqrt{s_{NN}} = 130$ GeV [1], are negative implying the correlation term $R_{+-}$ is larger than the combined $R_{++}+R_{--}$ terms. STAR moreover found $\nu_{+-,dyn}$ exhibits a qualitative "1/N" , with N being the number of particles produced by the collisions in the given (pseudo) rapidity range. The 1/N scaling was however found to be approximate only, given the scaled fluctuations $dN/d\eta\ \nu_{+-,dyn}$ exhibits a finite collision centrality dependence.

We begin our discussion of new analyses with a comparison, presented in Fig 1, of the energy dependence of the net charge dynamic fluctuations of the centrality dependence of $\nu_{+-,dyn}$ measured at $\sqrt{s_{NN}} = 20, 130,$ and 200 GeV.

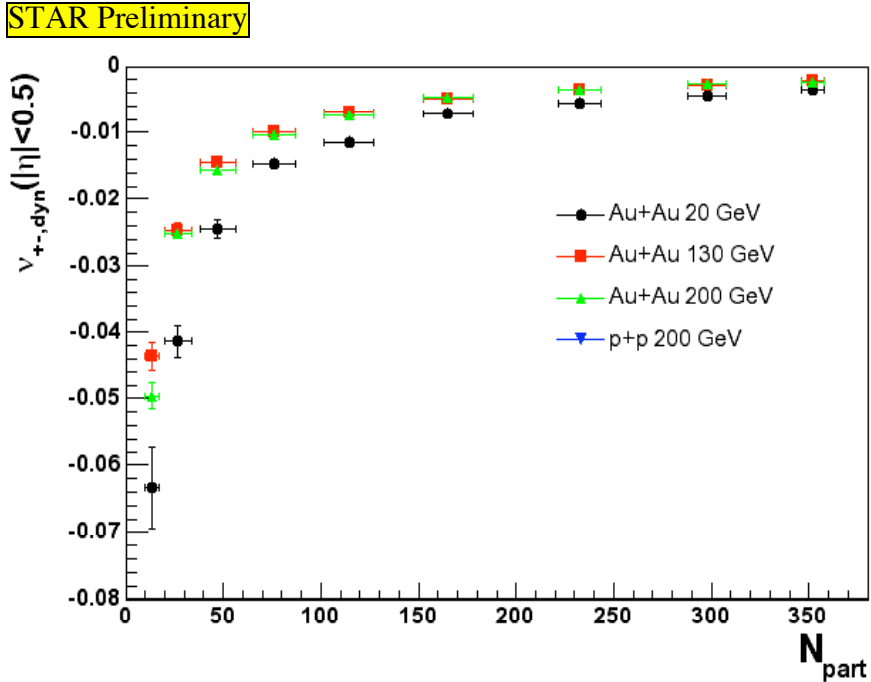

Fig. 1 Net charge dynamical fluctuations, $\nu_{+-,dyn}$, of particles produced with pseudo-rapidity |η|<0.5, as function of the number of participating nucleons.

Dynamical net charge fluctuations are finite at all three energy and exhibit a similar monotonic dependence with the number of participating nucleons. One finds net charge dynamic fluctuations measured at 130 and 200 GeV have a rather similar strength but are considerably smaller in magnitude than values measured at 20 GeV. It is known that a finite fraction of the $\nu_{+-,dyn}$ strength arises from charge conservation. At 130 GeV, this fraction was estimated in [1] to amount to 26% of the measured dynamic fluctuations in central collisions. Given the charge conservation contribution is inversely proportional to the total charged particle multiplicity, we therefore expect the larger $\nu_{+-,dyn}$ values observed at 20 GeV to arise mainly from charge conservation. A quantitative verification of this statement however awaits measurements of the total charged particle multiplicity at that energy.

The observed monotonic reduction of the magnitude of $\nu_{+-,dyn}$ arises principally from the progressive dilution of the correlation function when the number of particle production sources is increased. One expects $\nu_{+-,dyn}$ to be strictly inversely proportional to the number of participating nucleons or the produced particle multiplicity *if* the Au + Au collisions actually involve mutually independent nucleon-nucleon interactions, and rescattering effects may be neglected. We investigate the possibility of such a scenario by plotting the dynamical fluctuations scaled by the number of participating nucleons, the number of binary collisions, and the produced particle multiplicity (dN/dη) in Figure 2.

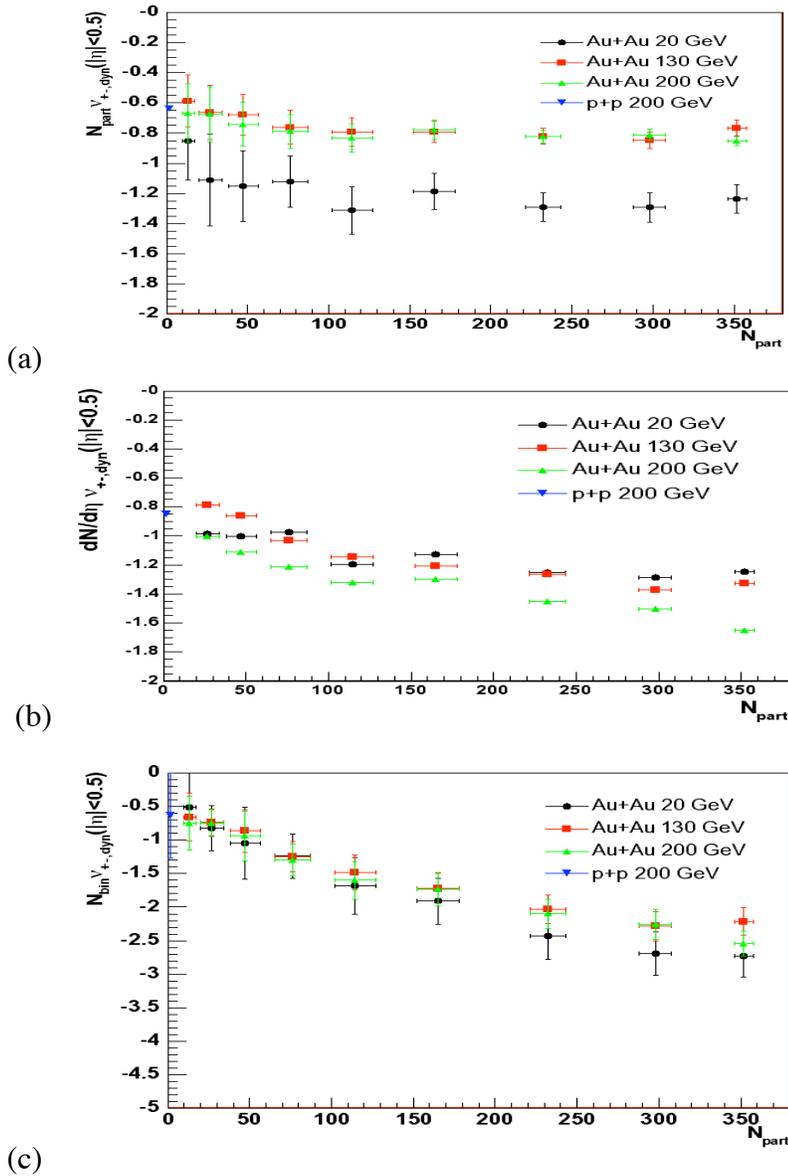

(a)
(b)
(c)
Fig 2 Net charge dynamical fluctuations, $\nu_{+-,dyn}$, of particles produced with pseudorapidity |η|<0.5 scaled by (a) the number of participants, (b) the produced multiplicity dN/dη, and (c) the number of binary collisions as function of the number of participating nucleons.

In Fig 2 (a), one observes that the scaled dynamical fluctuations, $N_{part} \nu_{+-,dyn}$, measured at 130 and 200 GeV are rather similar both in magnitude and collision centrality dependence, while they differ appreciably in magnitude from the 20 GeV values. Note however, once again, that such a difference may arise in part from differences in charged particle multiplicity production in nucleon – nucleon collisions at these three energies. It is nonetheless remarkable to notice that $N_{part} \nu_{+-,dyn}$, exhibit a finite centrality dependence at all three energies. Note also that the 200 GeV Au + Au measured in most peripheral collisions are in excellent agreement with p + p data measured at 200 GeV.

We investigate the dependence of produced multiplicity by plotting in Fig 2 (b) the dynamical fluctuations scaled by the produced multiplicity $dN/d\eta$. Note that values of $dN/d\eta$ used for this scaling correspond to efficiency corrected charged particle multiplicities previously measured by STAR[10,11]. We observe that all three distributions exhibit the same qualitative behavior: the amplitude $|dN/d\eta\, \nu_{+-,dyn}|$ is smallest for peripheral collisions, and rise monotonically by up to 50% in central collisions. We also find that the measured centrality dependence exhibit minor differences at the three energies, with the 20 GeV curve being the flattest.

As a matter of curiosity, we show in Fig 2 (c) the dynamical fluctuations $\nu_{+-,dyn}$ scaled by the number of binary collisions estimated on the basis of a Glauber Monte Carlo calculation [7]. We find, interestingly, that the three curves lie, more or less, on a "universal curve", and exhibit rather large collision centrality dependence. The significance of this "universal curve" is still subject to interpretation.

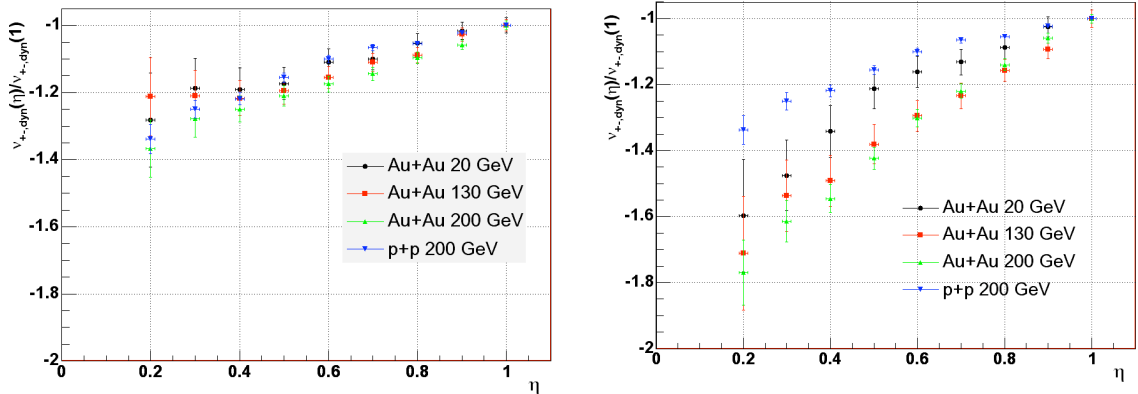

Figure 3: Dynamical fluctuations $\nu_{+-,dyn}$, normalized to their value at $\eta=1$, as function of the integrated pseudorapidity range, for Au + Au collisions at $\sqrt{s_{NN}}$=20, 130, 200 GeV, and p + p interactions at $\sqrt{s_{NN}}$=200 GeV for Au + Au centralities corresponding to 50-70% of the total cross section (left), and 0-10% (right).

The decidedly finite centrality observed in the scaled dynamical fluctuations presented in Fig 2 indicate that two-particle correlations are subject to a dramatic change in central collision relative to that observed in most peripheral collisions. We investigate this further by examining the dependence of the dynamical fluctuations on the width of pseudorapidity range used in the measurement. The measurements are conducted with a symmetric acceptance – $\eta_o < \eta < \eta_o$. The half width of the acceptance, $\eta_o$, is varied from 0.2 to 1, in steps of 0.1. Given the dynamical fluctuations at the three energies have rather different magnitude, we plot, in Fig 3, the dynamical fluctuations normalized to their value for $\eta_o=1$, as a function of the half width, $\eta$, of the integrated pseudorapidity range for two collision centrality bins. Au + Au data are plotted in the left panel for collisions corresponding to 50-70 % of the total cross section, and in the right panel for the 0-10 % most central collisions. Au + Au data are compared in both cases to inclusive p + p (i.e. integrated over all multiplicities). One finds that the magnitude of the normalized correlation is maximum for the smallest pseudorapidity ranges and decays monotonically to unity, at all energies and centralities, with increasing pseudorapidity range. The dynamical fluctuations being essentially a measure of two-particle correlation dominated by the $R_{+-}$ term, one finds, as expected, that the correlation is strongest for small rapidity intervals, and is increasingly diluted (reduced) for larger intervals. More interestingly, one observes that the magnitude of the correlations in central Au + Au collisions (right panel) increases monotonically with beam energy relative to those measured in p + p, whereas the magnitude of the correlations in peripheral Au + Au collisions (right panel) is essentially the same as that observed in p + p. The effect is rather dramatic if one considers that the relative magnitude of the correlations measured at $|\eta|<0.5$ increases by nearly 30 % in 200 GeV Au + Au relative to those in p + p. This behavior is, in essence, equivalent to reduction of the width of the balance function in central collisions relative to peripheral collisions reported earlier by STAR [12]. In central collisions, correlated pairs of negative/positive particles tend to be emitted much closer in rapidity than those in peripheral Au + Au or p + p collisions. Authors of Ref [13] have proposed that a reduction of the width of the balance function, and conversely a relative increase of short medium range ($\eta<0.5$) correlations, could signal delayed hadronization. The observed increased in the correlation, reported here, might however also result from the strong radial flow believed to exist in central Au + Au collisions. One speculates that measurements of the rapidity and azimuthal dependence of $v_{+-,dyn}$ and $p_t$ fluctuations (next section) might in fact permit a detailed study of the flow radial velocity profile produced in Au + Au collisions [14].

IV. Transverse Momentum Fluctuation Results

Measurements of finite non-statistical fluctuations of transverse momentum were reported for Au + Au at sqrt($s_{NN}$) = 130 GeV by both the STAR and PHENIX collaborations. In this work, we examine how these fluctuations evolve with beam energy and scale with collision centrality [3,4].

Figure 4 presents event-wise average transverse momentum histograms measured in Au + Au at $\sqrt{s_{NN}}$ = 20 and 200 GeV. Square and triangle symbols are used for real and mixed events respectively. Solid lines are a Gamma function fit to the data. A small but finite difference between the real and mixed events is manifest attesting that dynamical $p_t$ fluctuations are found at 20, and 200 GeV as well as at 130 GeV [3,4].

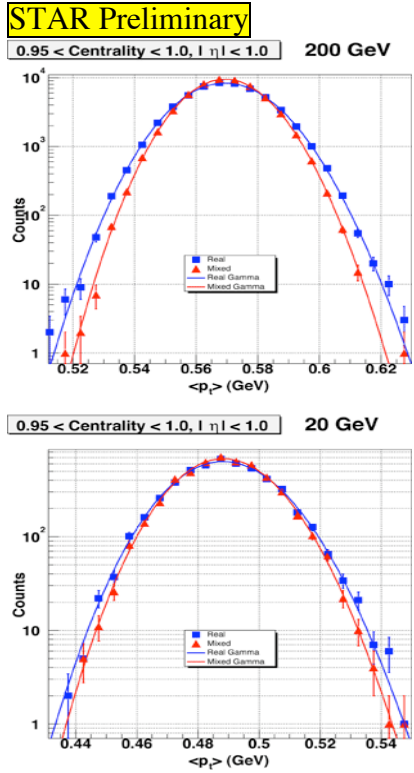

Figure 4 Histograms of event-wise average transverse momentum measured in 0-5 % most central Au + Au collisions at sqrt($s_{NN}$) = 20 (bottom) and 200 (top) GeV.

We proceed to investigate the behavior of these dynamical $p_t$ fluctuations using the average correlator <$\Delta p_{t,1} \Delta p_{t,2}$> defined in Section II.

Figure 5 presents our measurement of <$\Delta p_{t,1} \Delta p_{t,2}$> in Au + Au at sqrt($s_{NN}$) = 20, 130, and 200 GeV as function of the number of participating nucleons, $N_{part}$. One observes the correlator <$\Delta p_{t,1} \Delta p_{t,2}$> is finite at all three energies and that it exhibits a qualitative inverse proportionality to $N_{part}$. This qualitative dependence is known to arise from the progressive dilution of the correlation with increased number of particle sources. Experimental data (solid symbols) are compared to HIJING calculations shown with opened symbols. Although the HIJING calculations exhibit the same qualitative behavior with increasing collision centrality, one observe their magnitude is dramatically different from that of the measured correlations. Given this difference might in part result from an incorrect representation of two-particle correlations in HIJING, it is important to note that the ratio of experimental data to predictions vary dramatically from a value of 2.3 in peripheral collisions to a value of 4 in most central collisions. Clearly, a change in the collision dynamics arise in central collision relative to peripheral collision that is not predicted by HIJING.

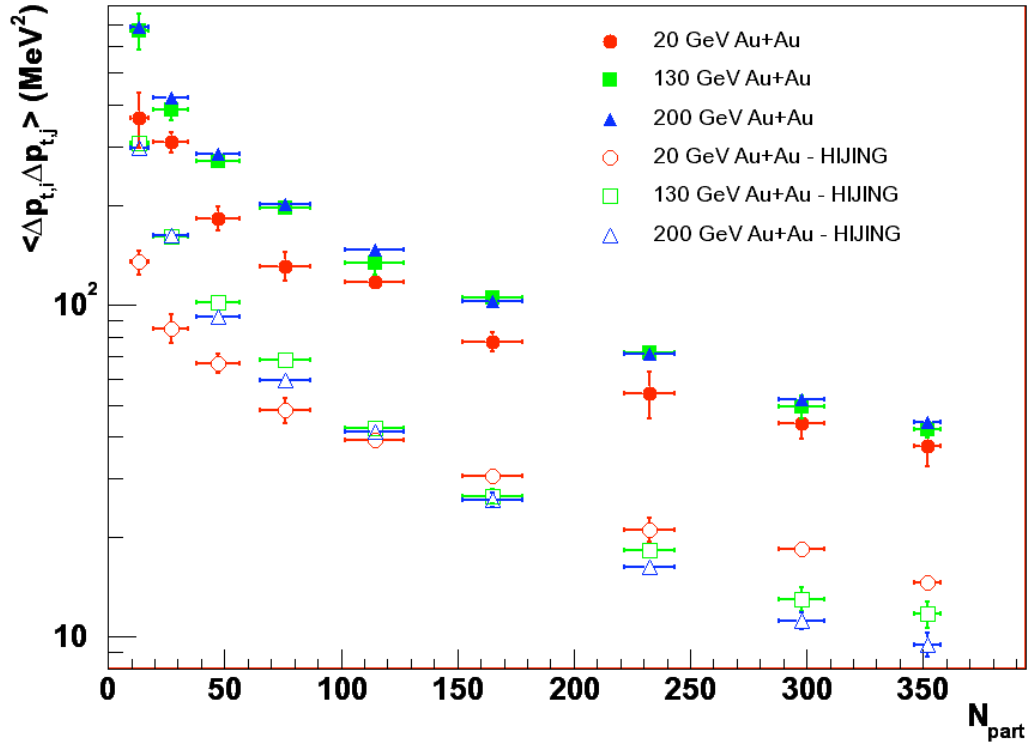

Figure 5  Comparison of $<\Delta p_{t,1} \Delta p_{t,2}>$ measured in Au + Au at sqrt(sNN)=20, 130, 200 GeV with prediction of HIJING plotted vs number of participating nucleons.

Another aspect of the measured correlation presented in Fig 5 is that the correlation $<\Delta p_{t,1} \Delta p_{t,2}>$ exhibits a small but nonetheless finite dependence on beam energy. Note however that $<\Delta p_{t,1}\Delta p_{t,2}>$ is observed to grow with beam energy whereas $\nu_{+-,dyn}$ was found to be larger at 20 GeV than at 130 and 200 GeV.  We speculate this different behavior may arise from increased radial flow going from 20 to 200 GeV, and show in Fig 6, a plot of the square root of the correlator $<\Delta p_{t,1} \Delta p_{t,2}>$ divided by the average transverse momentum as a function of the number of participating nucleons.  Data measured by the CERES collaboration in Pb + Pb collisions at 17 GeV are also included for comparison. We find STAR data, measured at the three energies, lie on a single curve. CERES data are in excellent agreement with STAR data in peripheral collisions but diverge slightly in most central collisions due possibly to differences in the experimental acceptance used by the two experiments.

We next study the monotonic decrease of the correlator $<\Delta p_{t,1} \Delta p_{t,2}>$ with increasing number of participating nucleons. Consider that if Au + Au collisions consisted of a superposition of independent nucleon-nucleon interactions, with no rescattering of secondaries, the correlator measured at a given centrality in Au + Au should be proportional to the correlator measured in p + p, and inversely proportional to the number nucleon-nucleon interactions at the given centrality. In such collision scenario, the produced particle multiplicity should be strictly proportional to the number of interactions. We thus scale the measured correlator $<\Delta p_{t,1} \Delta p_{t,2}>$ measured in the pseudorapidity range $|\eta|<0.5$ by the average number of produced particles $dN/d\eta$ measured in that range. Note that the quantity $<\Delta p_{t,1} \Delta p_{t,2}>$ is independent , similarly to $\nu_{+-,dyn}$, of detection efficiencies. Use of $dN/d\eta$ (corrected for detection finite efficiencies), rather then measured raw multiplicities, thus guarantees the product $dN/d\eta <\Delta p_{t,1} \Delta p_{t,2}>$ is properly corrected for efficiency dependencies on detector occupancies.

STAR Preliminary

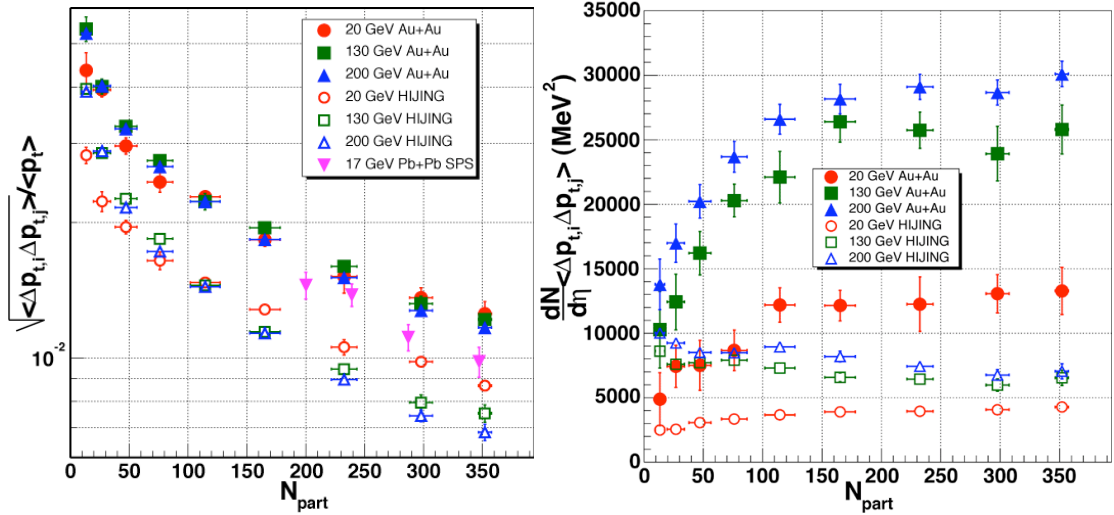

Figure 6 (a) Momentum correlation scaled by the inverse average transverse momentum as function of collision centrality. CERES Pb + Pb data at SPS shown in magenta.

Figure 6 (b) presents the scaled correlation $dN/d\eta <\Delta p_{t,1} \Delta p_{t,2}>$ as a function of the number of participating nucleons for the three measured energies (solid symbols). Also included in this plot are predictions based on HIJING. As a first observation, we note that the scaled correlator, $dN/d\eta <\Delta p_{t,1} \Delta p_{t,2}>$, exhibit a strong dependence on beam energy. Secondly, we note that, at variance with expectations based on an independent nucleon-nucleon collision scenario, the product $dN/d\eta <\Delta p_{t,1} \Delta p_{t,2}>$ also vary strongly with collision centrality at all three beam energies. Additionally, both these observations are at variance with the HIJING predictions, which show relatively small dependence both on beam energy and collision centrality. Violation of the "1/N" scaling expected for the independent nucleon-nucleon collision scenario is thus manifest: a dramatic change indeed occur in the collision dynamics of central Au + Au collisions relative to peripheral Au + Au and p + p collisions. This change may in part be connected to a build-up a radial collective flow believed to arise in Au +Au collisions. Gavin however argues that the observed behavior with centrality may also signal a large degree of thermalization is achieved in Au + Au collisions [15]. There exists also the possibility that the production of jets, and jet quenching may also influence those results. The rise of the jet production from 20 GeV to 200 GeV, included in the HIJING calculations presented in Fig 6, is however clearly insufficient to explain the much larger values found experimentally.

We investigate this problem further by studying the centrality dependence of the measured correlator $<\Delta p_{t,1} \Delta p_{t,2}>$ scaled by the number of participating nucleons, divided by the square of the average $p_t$, and multiply the ratio of multiplicities in p + p and Au + Au shown in Fig 7. The $N_{part}$ factor is used to account for the trivial dependence of the correlator on the number of sources of particles; the inverse factor $<p_t>^2$ should in part account for variations of the average transverse momentum with beam energy, and collision centrality; the ratio $(1+R_{AA})/(1+R_{pp})$ accounts for differences in two-particle correlations in Au + Au and p + p collisions.

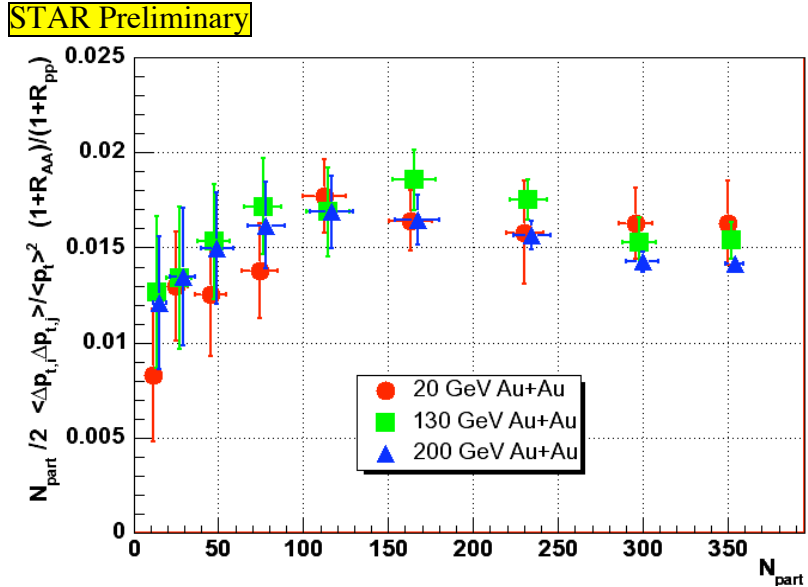

Figure 7 Measured correlator $<\Delta p_{t,1} \Delta p_{t,2}>$ scaled by number of participating nucleons, average transverse momentum, and relative multiplicities in p + p and Au + Au collisions.

We find the scaled data nearly lie of a single "universal" curve suggesting the performed scaling indeed eliminates "trivial" effects. Note however the scaled data nonetheless exhibit significant collision centrality dependence. We speculate this dependence arises mostly because of collective radial flow effects [14], and plan to investigate such effects further by studying the transverse momentum correlations dependence on integrated pseudorapidity and azimuthal ranges.

V.  Summary and Conclusions

We presented measurements of net charge and transverse momentum fluctuations in Au + Au collisions at $\sqrt{s_{NN}}$ = 20, 130, and 200 GeV using measures $\nu_{+-,dyn}$ and $<\Delta p_{t,1} \Delta p_{t,2}>$. We observed the dynamical fluctuations are finite at all three energies and exhibit a rather modest dependence on beam energy. We also found both net charge and $p_t$ fluctuations violate the trivial 1/N scaling expected for nuclear collisions consisting of independent nucleon-nucleon interactions. Scaled dynamical net charge fluctuations $|dN/d\eta\ \nu_{+-,dyn}|$ grow by up to 50% from peripheral to central collisions while $dN/d\eta\ <\Delta p_{t,1}\Delta p_{t,2}>$ exhibit an even larger change with centrality. We speculate the finite centrality dependence arise due in part to thermalization effects and in part to the large radial collective flow produced in Au + Au collisions. We therefore plan to extend this analysis by a study of the azimuthal dependence of $\nu_{+-,dyn}$ and $<\Delta p_{t,1} \Delta p_{t,2}>$ as well as a fully differential analysis of the two-particle correlations.